\documentclass{ws-ijmpe}

\begin{document}

\markboth{Haijiang Gong}{Search for Direct Photons from $\sqrt{s_{NN}}=200GeV$ AuAu Collisions Using a New $\pi^{0}$ Tagging Method in the PHENIX
Experiment at RHIC}

\catchline{}{}{}{}{}

\title{Search for Direct Photons from $\sqrt{s_{NN}}=200GeV$ AuAu Collisions\\
Using a New $\pi^{0}$ Tagging Method\\
in the PHENIX Experiment at RHIC
}

\author{Haijiang Gong for the PHENIX Collaboration}
\address{Department of Physics and Astronomy, State University of New York at Stony Brook\\
Stony Brook, New York 11794, USA\\
haijiang@skipper.physics.sunysb.edu}

\maketitle

\begin{history}
\received{(received date)}
\revised{(revised date)}
\end{history}

\begin{abstract}
Direct photons provide a insightful tool to study the different stages of a heavy
ion collision, especially the formation of a quark-gluon plasma, without
being influenced by the strong interaction and hadronization processes.
The yield of direct photons can be determined based on the inclusive
photon yield and the background from hadronic decays. We present a
new analysis technique applied to PHENIX Run4 Au+Au dataset.
It uses strict particle identification(PID) in the Electromagnetic Calorimeter(EMCal) and a charged particle
veto to extract a clean photon signal. These photons are then tagged with
EMCal photon candidates with loose PID cuts, which can be reconstructed with high efficiency, to
determine the fraction of photons originating from $\pi^{0}$ decays.
Many systematic uncertainties and detector effects cancel in this method.

\end{abstract}

\section{Introduction}
\label{introduction}

\begin{figure}[th]
\begin{minipage}[b]{0.45\linewidth} 
\centerline{\psfig{file=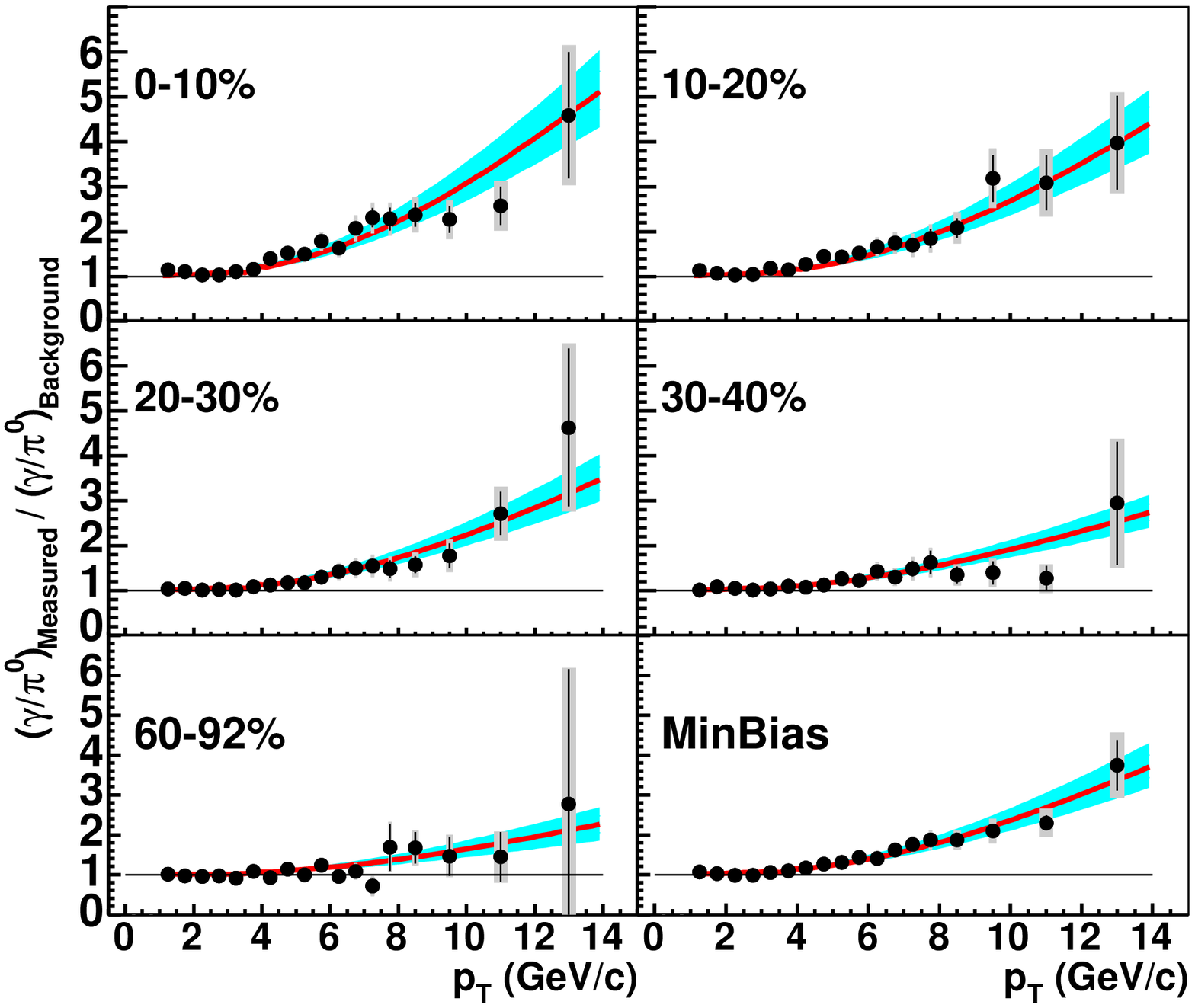,width=6cm}}
\vspace*{8pt}
\caption{Double ratio of measured $(\gamma/\pi^{0})_{Measured}$ invariant yield ratio to the background decay $(\gamma/\pi^{0})_{Background}$ ratio as
a function of $p_T$ for minimum bias and for five centralities of Au+Au collisions at $\sqrt{s_{NN}}=200GeV$(0-10\% is the most central). The solid
curves are the ratio of pQCD predictions described in the text.}
\label{publishresult}

\end{minipage}
\hspace{0.5cm} 
\begin{minipage}[b]{0.45\linewidth}
\centerline{\psfig{file=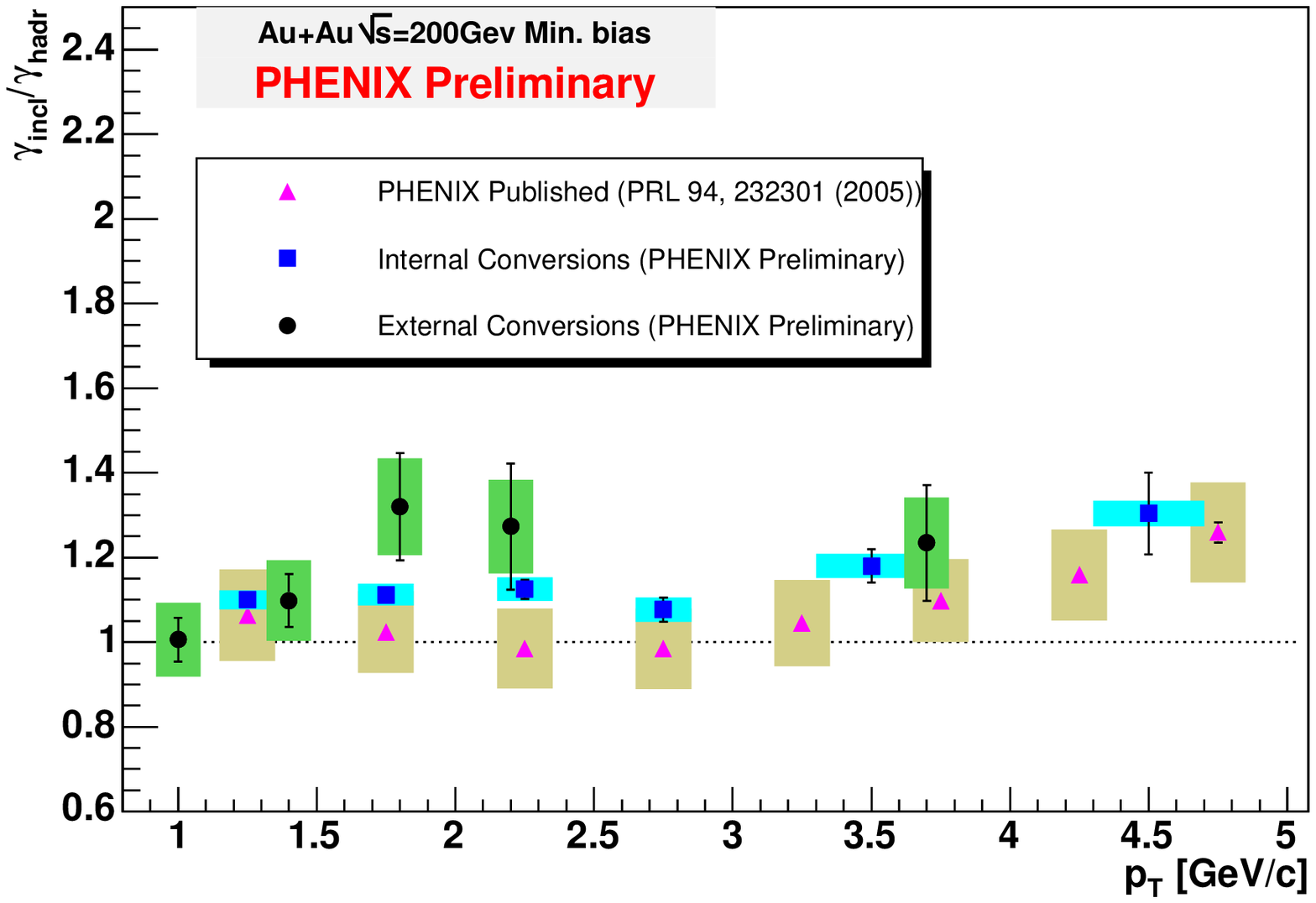,width=7.5cm}}
\vspace*{8pt}
\caption{Ratio of $\gamma_{inclusive}/\gamma_{hadronic}$ as a function of $p_T$ for minimum bias of Au+Au collisions at $\sqrt{s_{NN}}=200GeV$. The
purple triangles are the PHENIX publish result, the blue squares are from internal conversion method and the black dots are from external conversion
method. Color band on each data point is the systematic error and the extended line is the statistical error.}
\label{conversion}

\end{minipage}
\end{figure}

Recently PHENIX published measurement of direct photons in different centrality ranges in $\sqrt{s_{NN}}=200GeV$ Au+Au collisions up to
$p_T$=14GeV/c.\cite{1}At the high $p_T$ range, direct photons are produced in initial parton-parton scatterings, quark-gluon Compton scattering
$g+q\rightarrow\gamma+q$ and quark-antiquark annihilation $q+\overline{q}\rightarrow\gamma+g$ which are well described by perturbative Quantum
Chromodynamics(pQCD). The direct photon yield is determined through the comparison of the inclusive photon yield to the expected yield of background
photons from hadronic decays. The comparison is done in terms of the ratio of $N^{\gamma}/N^{\pi^0}$, or in short $\gamma/\pi^0$ since many systematic
errors such as energy scale uncertainties cancel. The photon excess then can be expressed directly as a double ratio of measured photons per $\pi^0$ to
the expectation from hadronic decays, $(\gamma/\pi^0)_{measured}/(\gamma/\pi^0)_{background}$. The double ratio in Fig.~\ref{publishresult} indicates
that there is a significant direct photons signal for $p_T\geq4GeV/c$ and it is consistent with the binary scaled pQCD model calculation from p+p
collisions.\cite{2} Apparently, the direct photons production is in direct contrast to high $p_T$ hadrons suppression observed in Au+Au
collisions.\cite{3}

At low $p_{T}$ range where a significant fraction of direct photons is expected to come from the thermalized medium of deconfined quarks and gluons,
the measurement is very challenging. These so-called thermal photons carry information about the initial temperature of the Quark Gluon Plasma(QGP).
But the huge background from $\pi^0$ decay and the relatively large systematic uncertainties at low $p_{T}$ limit our ability to make any statement
about this direct photon source.

In order to improve and understand the direct photon production in the low and medium $p_T$ range, PHENIX has already carried out several new analysis
techniques on RUN4 Au+Au $\sqrt{s_{NN}}=200GeV$ dataset, including using low mass dielectron pairs from internal conversions\cite{4} and external
photon conversions in the beam pipe.\cite{5} Unlike conventional direct photon method with the EMCal, these two new techniques both take the advantages
of the excellent capabilities of the PHENIX detector to measure electrons. Due to the excellent resolution of charged particles at low momenta, a
significant measurement for $1<p_T<5GeV/c$ was achieved(see Fig.~\ref{conversion}). The method presented here still uses photon candidates from EMCal
and conventional double ratio technique but instead of measuring $\pi^0$ yield in the double ratio of
$(\gamma/\pi^0)_{Measured}/(\gamma/\pi^0)_{Background}$, we measure $\gamma^{\pi^0}$ which are photons from $\pi^0\rightarrow\gamma\gamma$ decay by
reconstruction of photon pairs invariant mass. In this new method we will have to compare $\gamma/\gamma^{\pi^0}$ for inclusive photons in data with
the corresponding one for hadronic decay photons, which we can obtain from Monte-Carlo simulation. Since $\gamma$ and $\gamma^{\pi^0}$ are basically
the same physics quantities, many correction factors such as detector efficiency and acceptance cancel out explicitly and so do the systematic
uncertainties introduced by these factors.

\section{Analysis}

The data presented in this proceeding were collected during the 2004 Au+Au $\sqrt{s_{NN}}=200GeV$ run of RHIC by the PHENIX experiment. The EMCal in
the two central arms ($|\eta|\leq0.35$) is used to obtain the raw inclusive photon candidate. The EMCal consists of two subsystems: six sectors of
lead-scintillator sandwich calorimeter(PbSc) and two sectors of lead-glass Cherenkov calorimeter(PbGl). In this analysis, we only use the PbSc part and
events with the vertex $|z_{vertex}|<30cm$ from the collision center. That leaves us about 700 million minimum bias events after various quality
assurances. A series of stringent PID cuts are applied on photon-like clusters in the EMCal based on the time-of-flight(TOF) and the shower
profile($\chi^2$). Furthermore, charged particle contamination is removed by associating clusters in the EMCal with charged hits in the pad chambers(PC3) positioned directly in front of EMCal so it provides a clean photon sample. A minimum $p_T$ cut of $0.2GeV/c$ is also applied.

\subsection{Decay Photons tagged by $\pi^0$}
\begin{figure}[th]
\centerline{\psfig{file=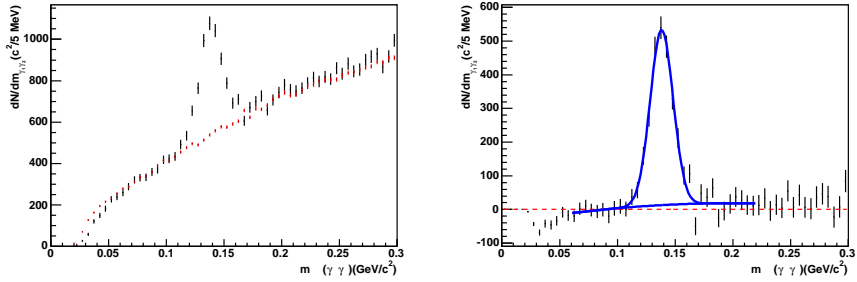,width=12cm}}
\vspace*{8pt}
\caption{Left panel: Invariant mass of tight-PID-cuts photons and loose-PID-cuts photons in same event and normalized mixed events for tight-PID-cuts
photons with $5.0<p_T<5.5GeV/c$. Black points are foreground and red points are normalized combinatorial background. Right panel: Invariant mass of
tight-PID-cuts photons and loose-PID-cuts photons after background substraction for tight-PID-cuts photons with $5.0<p_T<5.5GeV/c$. A fit of Gaussian
plus a second order polynomial is drawn.}
\label{peakextraction}
\end{figure}

In each event, these clean photons $N_{\gamma}$ are combined with photons reconstructed in the EMCal under loose PID cuts($\chi^2$ \& TOF \&
$p_T>0.2GeV/c$ but without PC3 veto) and their invariant mass is calculated, generating a foreground of physics pairs. The combinatorial background is
removed by event-mixing technique, which creates uncorrelated photon pairs from different events. The mixed event spectra is normalized to the
corresponding real photon pair invariant mass distributions below($50-100MeV/c^2$) and above($180-300MeV/c^2$) the $\pi^0$ peak then subtracted from
the foreground. Fig.~\ref{peakextraction} shows normalized background together with foreground(left) and subtracted foreground(right) for clean photon
$p_T$ range from $5.0GeV/c$ to $5.5GeV/c$. A Gaussian plus second-order polynomial function is used here to fit the $\pi^0$ mass peak since there is
still some background residual left beneath the peak after subtraction. So the contribution from $\pi^0\rightarrow\gamma\gamma$ tagged as
$N_{\gamma^{\pi^{0}}}$ can be determined by interpolating the Gaussian fit.
\noindent
\begin{eqnarray}
\label{eq:inclusive}
N_{\gamma}(p_T)=\epsilon\times\alpha\times N_{\gamma}^{incl}(p_T)/(1-X_{hadron})\\
\label{eq:piontag}
N_{\gamma^{\pi^0}}(p_T)=\epsilon\times\alpha\times\epsilon_L\times f\times (1-p_{conv})\times N_{\gamma^{\pi^0}}^{incl}(p_T)
\end{eqnarray}

The number of clean photons and the contribution tagged as coming from $\pi^0$ decay are measured as a function of $p_T$ of clean photons. As
Eq.~\ref{eq:inclusive} indicates, the measured yield of clean photons $N_{\gamma}$ depends on the reconstruction efficiency $\epsilon$ and the geometry
acceptance $\alpha$. Since charged tracks are removed by the PC3 veto cut, a hadron contamination correction mostly coming from neutron and antineutron
has to be applied where $X_{hadron}$ is the ratio of the remaining hadrons to all clusters that satisfy the clean photon's PID cuts. The tagged photon
yield $N_{\gamma^{\pi^0}}$(Eq.~\ref{eq:piontag}) has an additional dependence on the loose-PID-cuts photons' efficiency $\epsilon_{L}$, the conversion
correction $p_{cov}$ and the conditional probability $f$ to find a loose-PID-cuts photon in the PHENIX acceptance, given that its $\pi^0$ partner clean
photon is already reconstructed and accepted(see Section \ref{simulation}). Therefore, by calculating the ratio between $N_{\gamma}$ and
$N_{\gamma^{\pi^0}}$, the efficiency and acceptance for clean photons are explicitly canceled out.

The loss of $N_{\gamma^{\pi^0}}$ due to photon conversion $\gamma\rightarrow e^{+}e^{-}$ is corrected and the conversion correction is estimated by the
material budget between the collision vertex and the PC3 in front of the EMCal. The conversion material in RUN 4 has been estimated around $5-7\%/X_0$
and it is slightly different among different detector areas. Since we use a tagging method to measure clean photons from $\pi^0$ decay, only
loose-PID-cuts photons are affected due to the conversion loss. Therefore a correction factor of $(1-p_{conv})\sim94\%$ is applied.

\subsection{Simulation}
\label{simulation}
A similar ratio between the hadronic decay photon yield $N_{\gamma^{hadron}}$(Eq.~\ref{eq:calculated}) and the tagged photon yield from $\pi^0$ decays
$N_{\gamma^{\pi^0}}$(Eq.~\ref{eq:calculatedpiontag}) can be calculated from simulations where the acceptance $\alpha$ is canceled again.

\noindent
\begin{eqnarray}
\label{eq:calculated}
N_{\gamma^{hadron}}(p_T)=\alpha\times N_{\gamma^{hadron}}^{incl}(p_T)\\
\label{eq:calculatedpiontag}
N_{\gamma^{\pi^0}}(p_T)=\alpha\times f\times N_{\gamma^{\pi^0}}^{incl}(p_T)
\end{eqnarray}

The decay photon calculations are determined by a fast Monte-Carlo simulation of $\pi^0$ and $\eta$ decays. The PHENIX measured $\pi^0$
spectrum\cite{6} is used and the $\eta$ distribution is calculated based on $m_T$-scaling. The normalization factor of $\eta/\pi^0=0.45\pm0.10$ at high
$p_T$ is used.\cite{7,8}The EMCal energy resolution of $\delta_E/E=5\%\oplus9\%/\sqrt{E}$ is applied and a detailed description of the EMCal active
areas from data are used in the simulation. From the Eq.~(\ref{eq:calculated}) and Eq.~(\ref{eq:calculatedpiontag}) we will have
\noindent
\begin{equation}
(N_{\gamma^{hadron}}/N_{\gamma^{\pi^0}})_{calculated}=(1/f)\times
(N_{\gamma^{hadron}}^{incl}/N_{\gamma^{\pi^0}}^{incl})
\label{simulationratio}
\end{equation}
On the left side of Eq.~\ref{simulationratio} $(N_{\gamma^{hadron}}^{incl}/N_{\gamma^{\pi^0}}^{incl})$ is a straightforward physics quantity of excess
photons contribution from other hadronic sources than $\pi^0$ and depends only on the $\eta/\pi^{0}$ ratio and $\eta\rightarrow\gamma\gamma$ decay
branching ratio. So after applying the conditional probability $f$ which is depends on $\pi^0$ decay kinematics and the EMCal active areas, the ratio
$(N_{\gamma^{hadron}}/N_{\gamma^{\pi^0}})_{calculated}$ can be directly compared to the measured one.

In order to estimate the reconstruction efficiency $\epsilon_L$ of loose-PID-cuts photons, a PISA (PHENIX Integrated Simulation Application) simulation
of single particles with the complete PHENIX setup based on the GEANT package is conducted. The data from the simulation then is merged with the EMCal
data from the real events. By reconstructing the properties of these embedded particles, the occupancy effect of the detector in Au+Au collisions can
be estimated. After including all the loose-PID-cuts, the efficiency is determined to be around $82\%$ independent of $p_T$ beyond the minimum $p_T$
cutoff of $0.2GeV/c$.

The hadron contamination contribution $X_{hadron}$ is also calculated with a full PISA simulation. $\pi^0$, $\pi^{\pm}$, $K^{\pm}$ and $p\overline{p}$
are generated as input particles by using the actual PHENIX measured spectra.\cite{9}The spectrum of $n\overline{n}$ is estimated based on measured
proton and antiproton cross sections. For clean photons $N_{\gamma}$ since most of charged tracks are removed by the PC3 veto cut, the neutron and
antineutron contamination is dominant and it is largest around $p_T$ of $2GeV/c$ due to the contribution from annihilating antineutrons. Then it is
effectively reduced by the shower shape cut and becomes negligible above $4GeV/c$, which is less than $1\%$.

\section{Result}
The comparison of the ratio of measured $N_{\gamma}/N_{\gamma^{\pi^0}}$ and calculated $N_{\gamma^{hadron}}/N_{\gamma^{\pi^0}}$ is essentially the same
as the double ratio shown in Fig.~\ref{publishresult}. Any significant excess of the double ration beyond unity indicates a direct photon signal.
\noindent
\begin{equation}
R_{\gamma}=\frac{N_{\gamma}^{incl}}{N_{\gamma^{hadron}}^{incl}}=\epsilon_{L}\times(1-p_{conv})\times(1-X_{hadron})\times\frac{
(N_{\gamma}/N_{\gamma^{\pi^0}})_{measured}}{(N_{\gamma^{hadron}}/N_{\gamma^{\pi^0}})_{calculated}}
\label{ratio}
\end{equation}

\begin{figure}[th]
\centerline{\psfig{file=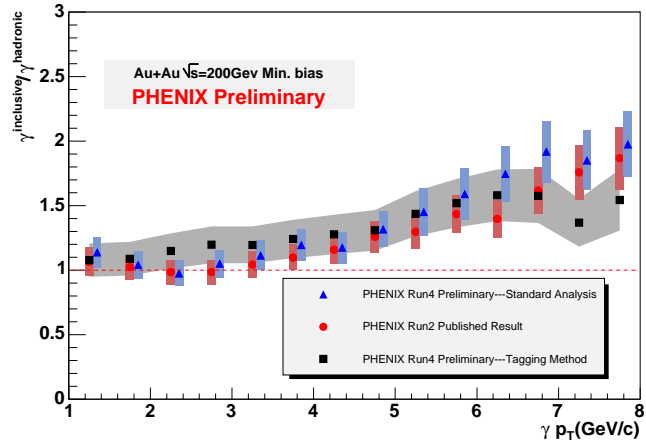,width=9.5cm}}
\vspace*{8pt}
\caption{Double ratio $R_{\gamma}$ as a function of photon $p_T$ for minimum bias in Au+Au $\sqrt{s_{NN}}=200GeV$ collisions. The solid black squares
are RUN4 tagging method result, the blue triangles are RUN4 conventional method result(points shifted) and the solid red dots are RUN2 PHENIX published
result. The color band around data points is the combined error.}
\label{finalratio}
\end{figure}

The preliminary RUN4 Au+Au results, in Fig.~\ref{finalratio}, show minimum bias double ratio $R_{\gamma}$ by using this new $\pi^0$ tagging method
compared with RUN2 published result and RUN4 preliminary result from conventional method. The main sources of systematic errors are the uncertainties
in the $N_{\gamma^{\pi^0}}$ yield extraction and the hadron contamination correction $X_{hadron}$, which gives a final systematic error on the double
ratio at the level of $10\%$. All results agree with each other within the error bar. For $p_T>2GeV/c$, the tagging method result offers a significant
direct photon measurement above 1. But for $p_T$ below $2GeV/c$ it is still limited by systematic errors. Compared with conversion methods mentioned in
Section~\ref{introduction} it has smaller statistical error and has the ability to extend to mid $p_T$ range.

\section{Summary}
With the extended statistics provided by the RHIC 2004 Au+Au run and this new powerful techniques to extract direct photon in medium and low $p_T$
range, a significant direct photon signal can be measured and results agree with PHENIX published result and several other new independent methods
including internal and external conversion. These promising measurement with improved systematic uncertainties indicates that for $p_T<5GeV/c$ direct
photon yield is lying above NLO pQCD expectation and suggests a possible thermal photon emission.


\end{document}